# Nuclear spin-density wave theory

Y. Cheng

*Department of Engineering Physics, Tsinghua University, Beijing, 100084, China*

Abstract

Recently [arXiv:0906.5417], we reported a quantum phase transition of $^{103m}$Rh excited by bremsstrahlung pumping. The long-lived Mössbauer excitation is delocalized as a neutral quasiparticle carrying a spin current. This letter gives a general theory for a nuclear spin-density wave propagating on crystals consisting of identical nuclei with a multipolar transition.

## 1. Introduction

Research in our laboratory has demonstrated the extraordinarily long-lived Mössbauer effect with a quantum phase transition at room temperature upon bremsstrahlung pumping and at 77 K upon cooling with liquid nitrogen [1-4]. Pumping the $^{103}$Rh nuclei from the ground state directly to the low-lying $^{103m}$Rh state is forbidden. Instead, the bremsstrahlung first pumps Rh nuclei to upper nuclear levels and then the nuclei cascade down to the long-lived $^{103m}$Rh state, as shown in figure 1(a). One of the major breakthroughs is the discovery of anomalous emission (AE) [5], which is observed not only from $^{103m}$Rh but also from the two long-lived Mössbauer isotopes $^{45m}$Sc and $^{93m}$Nb. The three nuclides are single isotopes with low-lying multipolar transitions. Cascade transitions occur without any intermediate eigenstate, as shown in figure 1(b), giving rise to an entangled biphoton [6, 7]. The biphoton propagation is transparent and recoilless due to strong coupling with the identical resonant nuclei in crystals.

The two coexisting massive biphoton modes are the polariton type and the plasmon type. The anisotropic mass of the polariton biphoton depends on the band structure, while the quasi-isotropic mass of the plasmon biphoton depends on its screening length. Analogous to the screened plasma mode, the screened plasmon biphoton is a short-ranged nuclear spin-density wave (NSDW) [4]. A similar concept, the so-called nuclear exciton [8], is traced back as an excitation wave in crystals to the Frenkel exciton [9]. The delocalized NSDW gives a better description for considering spin current, while the Wannier-Mott exciton leads to confusion with a Hydrogen-like atomic structure. This letter focuses on the $^{103m}$Rh NSDW, because concepts developed here are general and may be applied to the two isotopes $^{45m}$Sc and $^{93m}$Nb.

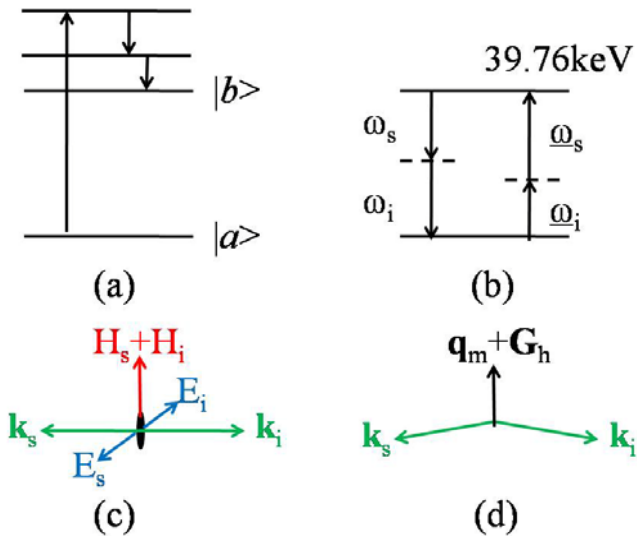

**Figures 1.** (color online) $|a\rangle$ is the $1/2^-$ ground state of $^{103}$Rh with odd parity. $|b\rangle$ is the $7/2^+$ excitation state of $^{103m}$Rh with even parity. The E3 transition from $|a\rangle$ to $|b\rangle$ is 39.76 keV with 56-min half-life. **(a):** Pumping path of the bremsstrahlung irradiation, where upper levels are firstly excited, then cascade to the $^{103m}$Rh Mössbauer state. **(b):** The cascade biphoton with frequencies $\omega_s$ and $\omega_i$ and the parametric biphoton absorption with frequencies $\underline{\omega}_s$ and $\underline{\omega}_i$. **(c):** One of the triplet biphoton modes with opposite momenta $\mathbf{k}_s$ and $\mathbf{k}_i$ simultaneously emitted from nucleus (shown by the black oval). The electric field $E_s$ and $E_i$ cancels each other; instead the magnetic field is enhanced. **(d):** $\mathbf{q}_m$ is the wave vector of a longitudinal acoustic phonon and $\mathbf{G}_h$ is a reciprocal lattice vector. The biphoton is the Borrmann mode with $\mathbf{q}_m = 0$, while it is the Brillouin scatterings with $\mathbf{q}_m \ne 0$.

## 2. NSDW concepts for the $^{103m}$Rh experimental results

AE from internal multipolar emitters deep inside a crystal has long been predicted by Hannon and Trammell [8], no experimental confirmation has yet been published. By matching the wavelength of the Mössbauer photon with the lattice constant, constructive interference at the Bragg angle leads to Hannon–Trammell's "AE", which is collectively enhanced by the presence of many Mössbauer resonators. However, the wavelength of the Mössbauer photon is on the nuclear scale, while the lattice constant is on the atomic scale. To match these two scales is almost impossible with a single Mössbauer photon. Instead, the scales are more naturally linked by the biphoton AE with its extra degree of freedom, as shown in figure 1(b).

The Borrmann mode, with a vanishing electric field at the lattice sites, is shown in figures 1(c) and 1(d). It decouples the electronic scatterings, and collectively couples the identical multipolar transition in crystals [8]. Borrmann emission is entirely recoilless, since the crystal as a whole takes the recoil momentum. Therefore, in the Borrmann mode, biphoton propagation is recoilless and is very weakly attenuated, making the crystal essentially transparent to this mode.

Although Brillouin scattering [10], shown in figure 1(d), is not the Borrmann mode, biphoton propagation remains recoilless and weakly attenuated by entering the strong-coupling regime [7, 9], where electronic scattering is also suppressed. The energy lost via a Stokes process is immediately recovered by an anti-Stokes process, conserving the energy and momentum by exchanging the biphoton with the phonon. In other words, the biphoton and the phonon emit and re-absorb simultaneously such that three quanta are entangled as the delocalized NSDW quasiparticle.

The two biphoton modes have different masses. The translational symmetry of the polariton biphoton is broken by Bragg reflection. The localized polariton biphoton acquires a mass on an order of keV [9]. The light mass (eV scale) of the NSDW manifests itself through a quantum phase transition at a critical density of $10^{12}$ cm$^{-3}$, reported elsewhere [4]. The mass is generated by the Anderson-Higgs mechanism [11], where the biphoton conspires with the longitudinal acoustic phonon.

The observed energy splitting, which is a function of the inversion density, is attributed to the polariton biphoton described by the Jaynes-Cummings model [7, 9]. The vacuum splitting of approximately 50 eV, which does not depend on the inversion density, appears in regime I, while the density-dependent splitting appears in regime II [1, 2]. Accordingly, the strong coupling opens a photonic gap with a persistent Rabi flopping. The splitting of K X-rays reveals that the atomic states are also dressed [4, 5] by the resonant coupling with the biphoton, since the energy of the individual biphoton is not fixed, as shown in figure 1(b). The vacuum splitting of 50 eV over the natural linewidth of $10^{-19}$ eV gives a collective amplification on the order of $10^{21}$, which increases to $10^{24}$ when taking into account internal conversion [12]. This tremendous amplification originates from the $N^2$ biphoton modes, where $N$ is the number of nuclei interacting. Conventionally, the splitting of the Jaynes-Cummings ladder is proportional to $n^{1/2}$, where $n$ is the number of photons in the cavity [7]. Instead, we observe a quadratic $n^2$ increase; i.e., a 9-fold increase in the energy splitting results from a 3-fold increase in density $n$ in regime II [2]. This crucial quadratic feature suggests a parametric 4-photon interaction.

Nuclei dressed by NSDWs short-cut the indirect pumping path, as shown in figure 1 (a), to give the observed nonlinear efficiency by means of parametric amplification [10] in regime II. As demonstrated by polariton-exciton Bose-Einstein condensation (BEC) [9], a straight BEC of the bosonic NSDW will not reduce the coupling. The spectral contraction in regime III indicates the abruptly reduced coupling that supports the Cooper paring of two NSDWs [4]. The nonlinear efficiency appearing in regime II is thus extinguished upon entering regime III. Pumping with more intensity leads to an earlier transition and a reduced NSDW density for a given pumping time [4].

At 300 K, the rotational symmetry of the NSDW giant magneton [4] is spontaneously broken, since the coupling strength is much larger than the thermal energy, as indicated by a Rabi splitting > 50 eV [1]. In fact, strong coupling and AE are generally observed in three regimes. Both translational and rotational symmetry of the NSDW are thus broken at an early stage in the bremsstrahlung irradiation. In other words, another phase transition for the diagonal long-ranged order will appear at low NSDW density, depending on the crystal perfection. Recently, Paredes and Cirac suggested the Cooper pairing of the fermionized 1D bosons [13], which may appear with 1D spin chains having NSDW texture, as previously discussed in [4].

## 3. Theory

The spontaneous decay of $^{103m}$Rh, with an internal conversion on the order of $10^3$ [12], exhibits a very weak coupling of $10^{-22}$ eV for the $^{103m}$Rh to free space transition. Due to the small volume of lattice cell, the biphoton coupling is estimated to be stronger than the coupling with a single photon. In the following formulation, we do not go into the detail of biphoton coupling but simply show that the strong biphoton coupling is due to the macroscopic number of the biphoton mode of $N^2$. The collective enhancement by $N \sim 10^{23}$ leads to the observable vacuum splitting on the order of 50 eV. We also estimate the mass and the size of the biphoton. The biphoton couples to the longitudinal phonon by Brillouin scattering under the phase matching conditions [10] given by

$$\begin{aligned} \omega_s + \omega_i &= \omega_M \mp \Omega_m \\ \mathbf{k}_s + \mathbf{k}_i &= \mathbf{G}_h \pm \mathbf{q}_m \end{aligned}, \quad (1)$$

where $\mathbf{q}_m$, $\Omega_m$ are the wavevector and the frequency of the acoustic phonon mode m, $\mathbf{G}_h$ is the reciprocal lattice vector of the Rh fcc crystal with the index $h$ corresponding to the plane $(h_1, h_2, h_3)$, $\omega_M$ is the Mössbauer transition frequency, and $\mathbf{k}_s$, $\mathbf{k}_i$, $\omega_s$, $\omega_i$ are wavevectors and frequencies for the signal and idler photons with subscripts $s$ and $i$.

The second line of (1) is the Bragg condition. The total energy of biphoton ($\mathbf{k}_s$, $\mathbf{k}_i$) is lost by emitting phonon $\mathbf{q}_m$ and vice versa. Since $\Omega_m$ is much less than $\omega_s$ and $\omega_i$, the phase matching conditions (1) imply that $\omega_s$ and $\omega_i$ are close to $\omega_M/2$ for the configurations shown in figure 2(a). Conversely, as shown in figure 2(b), the two frequencies differ when $\omega_s - \omega_i \sim cG_h$, where c is the light speed in vacuum. AE centered at half the transition energy of 19.9 keV is also attributed to the situation depicted in figure 2(a), whereas the two AE peaks of 17.4 and 22.4 keV are attributed to the situation depicted in figure 2(b). We have observed these three AE peaks and confirm that they exhibit a strong anisotropy for the configurations A and B, as shown in Ref. [4], which we will report elsewhere.

The superposition of the entangled biphoton is written as

$$\begin{aligned} \psi_{jkh}(\mathbf{r}_s, \mathbf{r}_i) &= \frac{1}{2\sqrt{N}} \sum_{\mathbf{k}_s, \mathbf{k}_i} \sum_{m=1}^{N} A_{jmh} A_{kmh} \delta(\omega_s + \omega_i - \omega_M \pm \Omega_m) \delta(\mathbf{k}_s + \mathbf{k}_i \mp \mathbf{q}_m - \mathbf{G}_h) e^{-i\mathbf{k}_s \cdot \mathbf{r}_s} e^{-i\mathbf{k}_i \cdot \mathbf{r}_i} \\ &= \frac{1}{2\sqrt{N}} \sum_{\mathbf{k}_s, \mathbf{k}_i} \sum_{m=1}^{N} A_{jmh} A_{kmh} \delta(\omega_s + \omega_i - \omega_M \pm \Omega_m) \delta(\mathbf{K} \mp \mathbf{q}_m - \mathbf{G}_h) e^{-i\mathbf{K} \cdot \mathbf{R}} e^{-i\mathbf{k} \cdot \mathbf{r}} \end{aligned}, \quad (2)$$

where $\mathbf{r}_s$ and $\mathbf{r}_i$ indicate the positions of the signal and idler photons, $\delta$ is the Kronecker delta function, $A_{jmh}$ and $A_{kmh}$ are the field components for the biphoton coupling with the phonon modes $\mathbf{q}_m + \mathbf{G}_h$, $N$ is the total number of coherent interacting nuclei, and $(\mathbf{r}_s + \mathbf{r}_i) / 2 = \mathbf{R}$, $\mathbf{K} = \mathbf{k}_s + \mathbf{k}_i$, $\mathbf{r}_s - \mathbf{r}_i = \mathbf{r}$, $\mathbf{k} = (\mathbf{k}_s - \mathbf{k}_i) / 2$. The low-energy dispersion relation of the longitudinal acoustic phonon [10] is $\Omega_m = vq_m$, where v is the sound velocity. The plus and minus signs before $\Omega_m$ in (2) represent Stokes and anti-Stokes scattering, respectively [10]. In the strong-coupling regime [7, 9], the energy loss by Stokes scattering will immediately be recovered by anti-Stokes scattering before the phonon disperses. Energy and the momentum are thus conserved by a superposition of the biphoton and the phonon in (2).

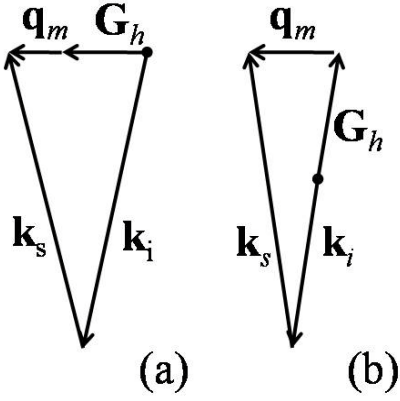

**Figure. 2.** (a) The biphoton with nearly equal energies
(b) The biphoton with different energies

Boson symmetry demands two possible wavefunctions for (2), of which the wave packets for the singlet biphoton $\psi_h^s$ with $s = 0$ or the triplet biphoton $\psi_{jkh}^t$ with $s = 2$ are written as

$$\psi_h^s(\mathbf{R},\mathbf{r}) = \frac{e^{i\mathbf{G}_h \cdot \mathbf{R}}}{\sqrt{N}} \sum_{m=1}^{N} \sum_j A_{jmh}^2 e^{i\mathbf{q}_m \cdot \mathbf{R}} \sin(\mathbf{k} \cdot \mathbf{r})$$

$$\psi_{jkh}^t(\mathbf{R},\mathbf{r}) = \frac{e^{i\mathbf{G}_h \cdot \mathbf{R}}}{\sqrt{N}} \sum_{m=1}^{N} A_{jmh} A_{kmh} e^{i\mathbf{q}_m \cdot \mathbf{R}} \cos(\mathbf{k} \cdot \mathbf{r})$$

(3)

The pseudo-scalar singlet of odd parity is irrelevant here due to its negligible coupling with the dark $^{103m}$Rh M4 transition. The transverse phonon modes do not couple to the triplet biphoton, whereas the longitudinal phonon modes, as the index waves, do couple to the biphoton. Due to the small nuclear radius $r$ with $kr \ll 1$, the collective coupling of the triplet biphoton is written as

$$g_h = \frac{1}{N} \sum_{n=1}^{N} \sum_{m=1}^{N} \sum_{jkl} \langle b, \mathbf{R}_n | O_{jkl} A_{jmh} A_{kmh} (\mathbf{q}_m + \mathbf{G}_h) \cdot \mathbf{e}_l | a, \mathbf{R}_n \rangle.$$

(4)

Here, $|a, \mathbf{R}_n\rangle$ and $|b, \mathbf{R}_n\rangle$ are the ground and excited states of the two-level nuclei at location $\mathbf{R}_n$, $\mathbf{e}_l$ are the unit vectors of coordinates and $O_{jkl}$ is the electric octupole of the E3 transition. The coupling $g_h$ is proportional to the macroscopic number $N$, which gives rise to an observable vacuum splitting of approximately 50 eV, as measured in regime I [1, 2]. Increasing the NSDW density $n$ in regime II, the Mollow-like triplet splitting [9] increases with $n^2$, as discussed in section 2, instead of with $n$ as revealed by the two-photon parametric amplification in (4). A higher-order approach is to replace the bare nuclei in (4) with the dressed nuclei in interaction. This polariton-type biphoton mode gives a heavy mass of approximately 2.3 keV for $K_\perp \to 0$ near the band edge with $\mathbf{K} = \mathbf{G}_{100} + \mathbf{K}_\perp$, which is approximated by setting the unknown reflective index $n_c$ = 1 with $\omega = (c/n_c)\sqrt{K_\perp^2 + G_{100}^2} \approx cG_{100}/n_c + cK_\perp^2/2n_c G_{100}$ [9]. Hence the reported phase transition is not attributed to this heavy polariton biphoton.

Instead, we suggest the massive NSDW of the triplet biphoton (s = 2, J = 3) is created by the Anderson-Higgs mechanism [11]. Here, the phonon is the Nambu-Goldstone mode due to the broken translational symmetry as a Higgs field. The plasmon biphoton acquires a mass of approximately 1 eV by absorbing the acoustic phonon. This plasmon biphoton can be interpreted as the $^{103m}$Rh NSDW, of which the screening length $\lambda \sim 1$ μm is roughly estimated by the plasma frequency $c/\omega_p$, in analogy with the London penetration length in superconductivity [14], as

$$\lambda = \sqrt{m_{Rh} c^2 / 4\pi n_{Rh} Z^2 e^2},$$

(5)

where $m_{Rh}$ is the rhodium mass and $n_{Rh}$ is the atomic density, $Z$ is the atomic number and $e$ is the elementary charge.

This light NSDW mass gives rise to a giant magneton to form 1D spin chains connected by the preferred end-to-end coupling [14]. The rotational symmetry is spontaneously broken at a critical NSDW density, which is rather dilute and is not observed in our experiments. We anticipate the BEC transition for the massive biphoton with spin multiplicity of 7 appearing at a density of $2 \times 10^{11}$ cm$^{-3}$ at 300 K [14]. However, the strong interaction of the NSDW chains inhibits the BEC transition. Instead, at the critical density $10^{12}$ cm$^{-3}$ the triplet biphotons are paired into a macroscopic condensate. The remaining vacuum splitting is attributed to the unpaired triplet biphotons. This model suggests that the phonon becomes massive, too. Accordingly, the critical temperature of the superconducting niobium crystal containing sufficient excited $^{93m}$Nb is expected to change.

## 4. NSDW at low temperature

The superfluid state of the Rh single crystal may have been observed at nK temperatures without recognizing that the $^{103m}$Rh excitation is generated by the background radiation penetrating the sample in the Helsinki cryostat located underground [15, 16]. Though the negative Weiss temperature θ has been repeatedly measured to be about -1.8 nK, the Helsinki group did not observe the antiferromagnetic nuclear ordering, even at the ultra-low temperature of 100 pK. Moreover, the Larmor resonance probed by an ac excitation along the static field shows a double-peak structure, where a narrow negative peak appears atop the normal positive peak [16]. The amplitude of the negative peak decreases roughly with the square of the nuclear spin polarization. To resolve these puzzles, we suggest that the negative contribution is attributed to the NSDW condensate. Recent progress on the superfluid phase of the exciton-polariton predicts a spin-Meissner effect [9, 17]. Accordingly, under an applied field of 4 gauss at nK temperatures, the extremely dilute NSDW condensate is fully polarized with $J_z = 3$ parallel to the field [4]. The majority of $^{103}$Rh nuclei (negative g-factor) with $I_z = -1/2$ decouple from $^{103m}$Rh with nuclear spin $I_z = 7/2$ (positive g-factor) due to the forbidden $\Delta I_z = 4$ transition. For T > 0, flipping $\Delta N$ nuclei from $I_z = -1/2$ to $1/2$ absorbs energy, while also releasing energy proportional to $\Delta N^2$ by coupling with the NSDW paring condensate, giving rise to a quadratic dependence on spin polarization. The diamagnetism may also negatively shift the Weiss temperature by suppressing the susceptibility.

Similarly, the spin-Meissner effect may have been observed by another group working on a Sc single crystal [18, 19]. A susceptibility peaking at sub-μK nuclear temperatures was taken as an indication for nuclear spin ordering. Interestingly, an anomalous suppression of the susceptibility appeared at the lowest nuclear spin temperature. Moreover, they observed an increasing polarization during the demagnetization warm-up under an external field of 7 gauss [19]. The theory presented in the current letter suggests a rather different aspect; *i.e.*, these two anomalies are mimicked by the spin-Meissner effect [4] attributed to the $^{45m}$Sc NSDW condensate. We have no information of the g-factor sign of $^{45m}$Sc in Ref. [12]. The majority of $^{45}$Sc nuclei with $I_z = 7/2$ (positive g-factor) are decoupled from any possible $^{45m}$Sc states with $-3/2 \leq I_z \leq 1/2$, when Sc NSDW is fully polarized with $J = 2$ and $J_z = 2$ parallel to the field. Assuming a negative $^{45m}$Sc g-factor, a negative contribution of the susceptibility as a quadratic function of the spin polarization may interpret the anomalies, as discussed above for the Rh case. Their unpublished result in figure 99 of Ref. [20] shows two successive peaks of the susceptibility during warm-up that may reveal the spin-Meissner effect switching off as a function of the unstable background radiation.

**Acknowledgments**

The author thanks the discussions with Hong-Jian He on the theoretical formulations and with J. T. Tuoriniemi and H. Suzuki on the low-temperature NMR experiments of Rh and Sc single crystals, respectively. This work was supported by the NSFC grant 10675068.

References
[1]   Y. Cheng, B. Xia, Q.-X. Jin and C.P. Chen (2008) (preprint arXiv:0807.1231).
[2]   Y. Cheng, B. Xia and C.P. Chen, Phys. Scr. **79** (2009) 055703.
[3]   Y. Cheng, B. Xia, C.-X. Tang, Y.-N. Liu, Q.-X. Jin, Hyperfine Interactions **167** (2006) 833.
[4]   Y. Cheng, "Quantum phase transition of the $^{103m}$Rh spin-density wave", (preprint arXiv:0906.5417).
[5]   Y. Cheng and B. Xia, "Anomalous emissions of $^{103m}$Rh biphoton transitions", (preprint arXiv:0908.0628).
[6]   Y. H. Shih, IEEE Journal of selected Topics in Quantum Electronics **9**, (2003)1455.
[7]   M. Fox, Quantum Optics, Oxford University Press (2006).
[8]   J.P. Hannon and G.T. Trammell, Hyperfine Interactions **123/124** (1999) 127.
[9]   A. V. Kavokin, J. J. Baumberg, G. Malpuech, F. P. Laussy, Microcavities, Oxford University Press (2007).
[10]  R. W. Boyd, Nonlinear Optics, ACADEMIC PRESS (2003).

[11] A. Altland and B. Simon, Codensed Matter Field Theory, Cambridge University Press (2006).
[12] R. B. Firestone (ed.), Table of Isotopes, 8[th] Eds., John Wiley & Sons, New York, 1999.
[13] B. Paredes and J. I. Cirac, Phys. Rev. Lett. **90** (2003) 150402.
[14] A. J. Leggett, Quantum Liquids, Oxford University Press (2006).
[15] T. A. Knuuttila, J. T. Tuoriniemi, K. Lefmann, K. I. Jntunen, F. B. Rasmussen, and K. K. Nummia, J. Low Temp. Phys. **123** (2001) 65.
[16] P. J. Hakonen, R. T. Vuorinen, and J. E. Martikainen, Phys. Rev. Lett. **70** (1993) 2818.
[17] Y. G. Rubo, A. V. Kavokin, I. A. Shelykh, Phys. Lett. A **385** (2006) 227.
[18] Y. Koike, H. Suzuki, S. Abe, Y. Karaki, M. Kubota and H. Ishimoto, J. Low Temp. Phys. **101** (1995) 617.
[19] H. Suzuki, Y. Koike, Y. Karaki, M. Kubota, H. Ishimoto, in Proc. of LT21, Prag, Part S4 (1996) 2183.
[20] A. S. Oja and O. V. Lounasmaa, Rev. Mod. Phys. **69** (1997) 1.